\documentclass[twoside]{article}
\usepackage{graphicx}
\usepackage{balance}
\usepackage{float}
\usepackage{times}

\newcommand{\clearemptydoublepage}{\newpage{\pagestyle{empty}\cleardoublepage}}

\begin{document}

%Put the date below the TR number
\date{}

%The title, authors and TR number show through the window:
\title{\vspace{1cm}{\huge Mining the Temporal Evolution of the Android Bug Reporting
  Community\\via Sliding Windows}}

\author{ Feng Jiang \and Jiemin Wang \and Abram Hindle \and Mario A. Nascimento}

\maketitle
\thispagestyle{empty}

\begin{center}

\vspace{-0.5cm}
{\large }

\vspace{2cm}

{\large September 2013}

\vspace{4cm}
{\large
Department of Computing Science\\
University of Alberta\\
Edmonton, Alberta, Canada\\
}

\vspace{2cm}

\copyright \/ University of Alberta

\end{center}

\pagenumbering{Roman}
\clearemptydoublepage
\setcounter{page}{1}

\begin{centering}
\section*{Abstract}
\end{centering}
The open source development community consists of both paid and volunteer
developers as well as new and experienced users.
Previous work has applied social network analysis (SNA) to open source
communities and has demonstrated value in expertise discovery and
triaging. 
One problem with applying SNA directly to the data of the entire project lifetime is that the impact of local activities will be drowned out. In this paper we provide a method for aggregating, analyzing, and visualizing local (small time periods) interactions of bug reporting participants by
using the SNA to measure the betweeness centrality of these participants.
In particular we mined the Android bug repository by producing 
social networks from overlapping 30-day windows of bug reports, each sliding
over by day.
In this paper we define three patterns of participant
behaviour based on their local centrality. We propose a method of analyzing the centrality of bug
report participants both locally and globally, then we conduct a thorough case study of the bug reporters' activity within the Android
bug repository.
Furthermore, we validate the conclusions of our method by mining the Android version control system and inspecting the Android release history.
We found that windowed SNA analysis elicited local behaviour that were invisible during global analysis.

\clearemptydoublepage
\tableofcontents
\clearemptydoublepage

\pagenumbering{arabic}
\setcounter{page}{1}

\section{Introduction}
% two columns !!!

\label{introduction}
% no \IEEEPARstart

%XXX cite
% Introduce SNA
% Introduce bug repository
% Social network analysis (SNA) is a powerful tool that helps
% practioners and researchers study the complicated interactions of
% participants within communities, SNA analysis is well accepted within
% the software maintenance and mining software repositories communities.
% SNA allows us to study the structure of the interactions between software developers and users
% within a software project's community using graph metrics on the graph
% of discussions that occur in the bug repository.
% Since the interaction in the communities records the activities of
% each participant and the communication among them

%SNA
Global analysis provides us with easy to interpret data that gives us an overview of the entire system. It simplifies complicated dimensions like time and provides us with an easy way to explain results. Unfortunately, for tools like \emph{Social Network Analysis} (SNA), a global analysis can miss a lot of important interactions, especially between stakeholders, thus we propose a method of using SNA to study bug repositories and tease out local collaborations.

SNA is a powerful tool that helps
practitioners and researchers study the complicated interactions of
participants within communities; SNA is well accepted in the area of software maintenance and mining software repositories
communities~\cite{CAMB:wass,MSR:christ,ICSEsocio:meneely}.  
The bug repository records interactions among software developers and users in a software
project's community. With SNA, we are able to study the structure of the interactions by analysing the graph constructed through the interaction of bug reporters in the bug repository.
The results can be used in expertise elicitation and
triaging in order to suggest which participants have expertise relevant
to an issue~\cite{ICSEsocio:meneely}. 
Usually SNA is run globally across all day, over a single period, or over an entire project lifetime. In this paper we argue that using SNA in a more local manner provides valuable insights into interactions between stakeholders during the development and maintenance of a software system.

Open-source communities are amenable to social network analysis as
they are
open to user interaction and participation. 
At the same time there is a lack of imposed organizational
structures found within corporate organizations~\cite{ACM:chris}.
Because open source projects often lack strict centralized control and
requirements~\cite{AMCIS:Freeh}, developers often choose their
tasks instead of being assigned one~\cite{ACM:ashish}. 
This fact suggests that local structure of interactions among users and developers
who express an interest in one part of the project tend to self
organize and produce interesting collaboration structures (networks).

%Bug repos
Bug repositories are also amenable to social network analysis as bug
repositories host and record discussions regarding issues or bugs relevant to
the development and the use of a software development
project~\cite{ACM:ashish,OSD:yasu}. 
Bug repositories are also heavily used by open-source projects.
Collaboration among developers has been studied in various aspects about how the communication introduces or avoids bugs, and further influences the software quality, \cite{CSMR:bernardi}, \cite{ACM:abreu}, \cite{ACM:pinzger}, \cite{IEEE:bettenburg}. 
Besides the collaboration among developers, collaboration between users and developers is evident in bug
reports since the discussions and communications are recorded as reported bugs, and posted comments
on bug reports.
One point here is that, both users and developers are often periodic, and their activities or collaborations can be local and thus missed out in global analysis.

%In this paper, we studied the Android bug community. The Android development contains both the systems or the sdks and the apps; it covers varieties of technique topics, and at the same time, developers are not organized statically as the very standard commercial software. The Android development comprises participants from different vanders, for example, some are from Motorola, some are from HTC, or Samsung, etc, and thus it should be interesting and worthwhile to study the Android community and see how well it could relate to the development facts. 

% Bug participants
In the case of the Android bug repository, provided by the 2012 MSR
Mining Challenge~\cite{DATA:msr}, a reporter would report a bug, which
might attract comments from bug commenters; the commenters discuss the
reasons and possible resolution of the bug.
The bug reporting community members are usually comprised of both bug
reporters and bug commenters who are either Android developers or
Android users.
From the perspective of the bug repository, unlike the version control
system, there is actually no
obvious boundary between a user and a developer.
We refer to these different participants as \emph{bug participants}.

% Setup SNA
In order to apply SNA to the bug repository, we first create the graph based on the interactions. 
We pose that each node of the graph represents
one bug participant and each edge represents the connection between
two participants who have communicated on the same bug. 
We will introduce the network graphs in detail in
Section~\ref{methodology}.

% SNA Metric
We use betweenness centrality to quantify the importance of a
participant in the community \cite{ICSEsocio:la} 
(betweenness centrality will be better explained in 
 Section~\ref{methodology}).
The betweenness centrality could reveal two aspects of a participant
in a community network: 1) the quantity of bug reports (which attract
at least one comment) or comments they have made and 2) the importance
of the content of their reports or comments. 
When participants have high betweenness, they might have: 1) reported
quantities of bugs with at least one comment on them, 2) made lots of
comments, 3) reported a very critical bug which attracts comments, 4) or made a very
interesting comment which attracts comments from other participants.

% Previous work which applied SNA to software development has been
% introduced such as mining email archives and mining defect tracking
% systems \cite{MSR:christ} \cite{ACM:ashish} \cite{OSD:yasu}. 
% Similarly, we built our SNA on bug reporting repository, which
% includes both bug reports and comments, and looked into the Android
% bug community based on the participants' activities and discussions.

% But we use windows
However, the previous work~\cite{MSR:christ,ACM:ashish} applied 
SNA on the entire lifetime of a project, such that only a single community network was constructed. 
Some of collaborations might not be evident if one were to
analyze a large single network.
That is because certain structures will not be observable on the global scale.
In order to peer into these local self organized structures using
social network analysis, we felt it is better to choose a \emph{windowed} approach, \cite{ICSEsocio:meneely, ICSMwindowed:hindle}. 
Windowing allows us to look at network during a slice of time
and then relate our measures (betweenness centrality per author) to
the next window and beyond. 
This \emph{sliding window} view of centrality allows us to see those
developers and users who are constantly at the forefront of discussion or
those who ebb and flow between issues and tasks.
Moreover, by sliding windows, each pair of adjacent windows would have
an overlap, which results in smoother trends, and more importantly,
helps to maintain context. 
Other benefits provided by time windowed analysis is that it gives a more accurate and nuanced view of the data as 
locally central participants then will not be 
``drowned out''.

In summary, we use SNA to study the activities of bug
participants based on the Android bug reports and comments
repository. 
We apply the sliding window method to observe smooth change trends
in the collaboration graph across time.
With these mining results, we seek to analyze bug participants'
interactions, activity trends and patterns. 
We then demonstrate our analysis results via answering the following
research questions about local and global behaviours.

Global research questions:

\textbf{RQ1.} How does the number of active bug participants change over
  time? Why?

\textbf{RQ2.} How does the betweenness centrality of a participant change over
time? What are the reasons when they have a certain activity pattern?

Local research questions:

\textbf{RQ3.} Are there special time ranges during which participants are more/less
active or central than normal? Why?

\textbf{RQ4.} What are the possible scenarios for a very sharp change of the
participants' centrality? Why?

We also validate if this windowed methodology actually
highlights relevant behaviour 
by inspecting the Android release
history\footnote[1]{Android release history:
  http://developer.android.com/sdk/index.html} and the Android version
control system.
The validation would be discussed in Section
\ref{validation}.

The rest of our paper is organized as follows. 
Section \ref{background} introduces basic concepts and techniques we
used in this study. 
The specific steps and the methodology will be discussed in Section
\ref{methodology}. 
Section \ref{results} describes the details of our mining results. 
The analysis of the results and its corresponding validation is
provided in Section \ref{validation}. 
Section \ref{limitation} presents the limitations of our mining
process and Section \ref{conclusion} summarizes the paper and
discusses the future work.

%%%%%%%%%%%%%%4. What are the possible scenarios for the very sharp change in the participants' centrality?

% You must have at least 2 lines in the paragraph with the drop letter
% (should never be an issue)
\begin{figure}[!t]
\centerline{\includegraphics[width=2.6in,height = 3.8cm]{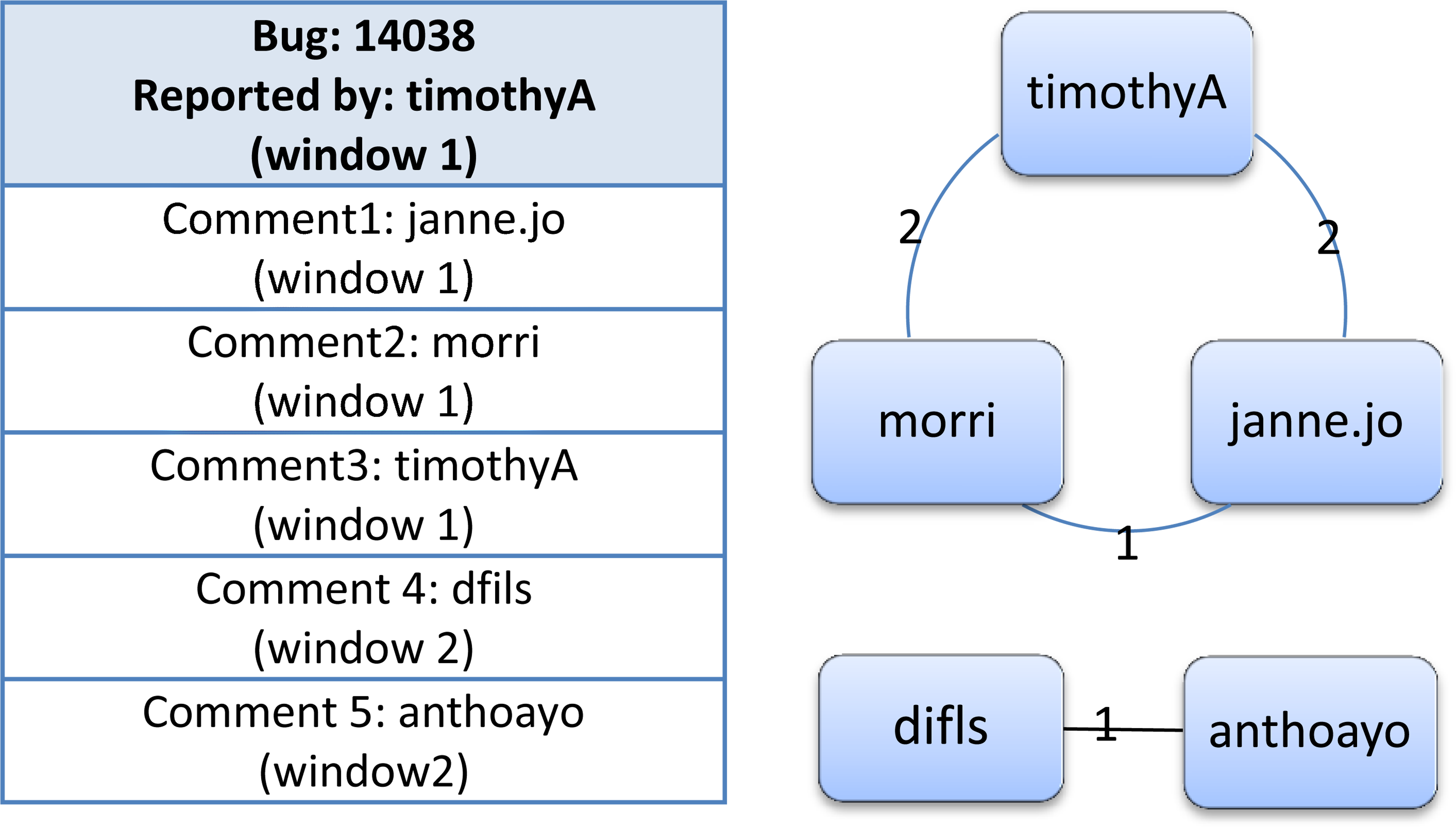}
\label{graph}}
\caption{An example: bug 14038 is reported by timothyA, and there are
  five comments on this bug. 
When time window applied, comments are plotted into two windows, and
the bug report of this example forms two networks with the weight
noted on their edges}

\end{figure}

\section{Background}
\label{background}
% talk about window, overlap, cluster 
% need to take up 2 columns
\subsection{Betweenness Centrality}

% XXX Defined by Freeman? Link/Cite!
The betweenness centrality of
a vertex is the number of geodesic paths in a graph that includes this
vertex;
the geodesic path is defined as the shortest path which has the
minimum weight between two nodes. 
Defined by Freeman \cite{SOCIO:freeman}, the betweenness can be represented as:

\begin{equation} 
\sum_{i=1}^{j-1}\sum_{j=1}^{n}\frac{g_{ij}(k)}{g_{ij}}, i\neq j \neq k
\end{equation}

where $k$ is a vertex of the graph, $n$ is the total number of
vertices, $i$ and $j$ are vertices other than $k$, $g_{ij}$ is the
number of geodesic paths between vertex $i$ and $j$, and $g_{ij}(k)$
is the number of geodesic paths that include $k$.

It is used as a measurement of a person's importance in a network. 
A person would be regarded as central if he is on the geodesic path
between two other persons. 
As proposed by Freeman~\cite{SOCIO:freeman}, if a person is located
on the geodesic path between two other persons, he becomes one of the key
persons who connects the others. 
That is, the more a person connects to
the other people in a network, the more important or central he is~\cite{BOOK:han}.

In our work, we normalize the betweenness centrality values to eliminate the effect of different sizes of the
networks. The betweenness is normalized as:

\begin{equation}
\label{normalizedb}
Normalized~B = \frac{B}{\frac{(n-1)(n-2)}{2}}
\end{equation} 

where $B$ represents the original betweenness value and $n$ is the number of nodes in the graph being calculated.

Compared with simply counting the total number of comments or total
bug reports of a participant, betweenness acts better to reflect the
interactions among people. For example, when a person reports lots of
bugs but none of them attract any comment, it is very likely that
his bug reports are not interesting or important. In this case, if
we merely counted the number of their reports or comments, we would
possibly increase their importance in the network
artificially. Therefore, we choose to use betweenness centrality to
eliminate this unfair counting~\cite{ICSEsocio:la}.

\subsection{Overlapping Time Windowing}
\label{overlappingtimewindowing}
When SNA is applied in other
papers~\cite{MSR:christ,ICSEsocio:meneely}, it is typically applied to
the entire history or one period of the partial history and all the
bug reports within that period.
Windowed analysis instead repeats social network analysis across 100s
of windows (in our case, as many windows as we have days). These
windows overlap and often the analysis of one window results in the
same analysis as the previous window due to the overlap. We slid our
windows by 1 day and for two adjacent windows
$A$ and $B$, $B$ starts on the second day of $A$, and they would have
an overlap of 29 days, that is each window does some redundant analysis but produces 
smoother transitions in
analysis between windows. Thus 1 comment in a bug report will have an
effect on the graphs of 30 windows. This is similar to Hindle et
al.'s~\cite{ICSMwindowed:hindle} analysis of topics using windows but
they did not use an overlap. We could thus see the changes in the trend of a participant's activity.

Moreover, time windowed analysis could give a more accurate and
nuanced view of the data \cite{ICSEsocio:meneely, ICSMwindowed:hindle}, as locally central participants would not
be ``drowned out''. For instance, if a
bug participant participates in many bug reports and bug comments during one
month, he would be one of the most central participants with a high
betweenness within this window. However, if he appeared only for that
month, globally, he would have low betweenness and would not show up
as central, even though during a shorter period he played a vital role. 
As we can see in Figure 2, the left column graph shows the betweenness values of participants 
over the entire time period; local details are missed and we get nothing about the trend, compared to the right part of the results from overlapping windowing. For example, cluster 8 on Figure 2 is bright and important at the start of our analysis but does not appear in the global graph on the left. Also, if there
is a very sharp drop of values of a certain participant, the
overlapping windows would give a more nuanced view of the change and
what was happening.

Another point is that, comments on the same bug might not be globally
temporally relevant~\cite{Springer:kidane,Procedia:ibaa} thus 
 a global time analysis would not make much sense in this case. This
 could happen if new changes induce new bugs or modify the behaviour
 of a reported bug.

\begin{figure*}[ht]
\centering
\includegraphics[width=\columnwidth]{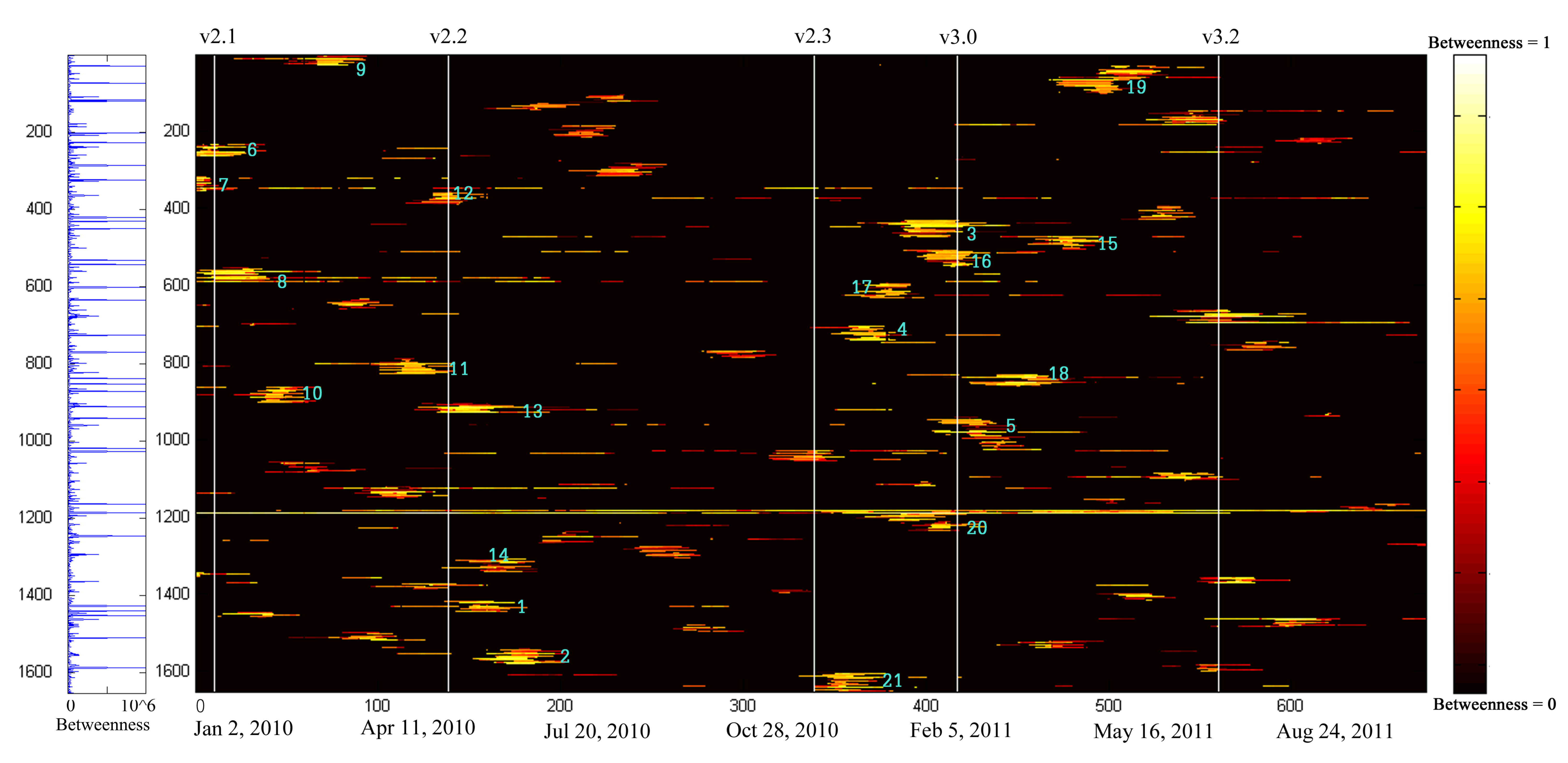}
\label{result}
\caption{Betweenness centrality along time line: the x-axis represents the
number of time windows and the starting dates are denoted every 100
windows. The y-axis represents the number of bug participants who have
ever been central in the bug community with betweenness
centrality valued greater than 0 for some time period. The
color represents the value of betweenness centrality, with darker
colors corresponding to lower betweenness and lighter colors for
higher betweenness. We used K-means clustering with cosine distance where K = 100 }
\end{figure*}

\subsection{Clustering}
Figure 3(1) orders participants by their betweenness values. We can indeed find that there are participants of low overall betweenness but being very active (show up bright) at some time points, and this supports the necessity of windowing, as stated in Section \ref{overlappingtimewindowing}. However, we need more information about participants' working patterns and get an idea about their being interacting groups. 

In order to perceive clusters, that are local groups of interactions,
we clustered the bug participants using K-means by their betweenness centrality
distribution along the time line.  
K-means is one of the most popular
clustering methods which aims to partition $n$ data items into $k$ clusters that each data item belongs to the cluster with the nearest mean \cite{MATH:macqueen}. 
We choose to use K-means with cosine distance.
The
cosine distance between two vectors is defined as,

\begin{equation}
cosine\_dist(A,B) = 1-\frac{A\cdot B}{\Arrowvert A \Arrowvert \Arrowvert B \Arrowvert}
\end{equation}

where A and B are two vectors, $\cdot$ represents the inner dot
operation and $\Arrowvert \cdot \Arrowvert$ indicates the module of
the vector. With clustering, authors with similar temporal centrality would be
grouped together so that bug participants with similar activity
patterns are also grouped together. 

In this paper, we choose the K-means with cosine distance because it gives a better visual clustering result, as compared in the plots of Figure 3.
In this case, cosine distance calculates the similarity between each pair of
participants in terms of temporal centrality whereas Euclidean distance
focuses on the magnitude of data, the size and frequency of centrality.

Moreover, we used $k = 100$ for the K-means, to cluster authors. There is a trade off between the size of clusters and the variance within each cluster. As we can see from Figure 3(3), $k = 10$ also gives good visualization result, but considering the number of more than 1600 participants, we should get a larger $k$ to keep the diversity of working groups in similar. We set $k = 100$ in this case, since we get the aesthetically best visualization (our
subjective opinion based on visible clusters) of all the data; we had tried other values of $k$ such as 5, 10, 25, 50, 75, 200, 300. 

\section{Methodology}
\label{methodology}
Our methodology consists of six steps that deal with raw data, construct graphs and apply social network analysis with sliding windows. We conduct a thorough case study of the Android bug repository with the proposed method and validate the conclusions from the results by mining the Android version control system and inspecting the release history.
\subsection{Data}

With the provided Android bug repository 2012 and the Android version
control system
from the MSR Challenge~\cite{DATA:msr}, we converted and stored the XML format data into 
a database for efficient analysis using Microsoft SQL Server Business
Intelligence. Our analysis focused on the bug records of the previous
two years from January 1st 2010 to December 4th 2011 since during these two years, there are more records in the repository as we counted that participants are more active; also, the
activities are representative, both the Android platforms and their developer groups are larger and more diverse during the latest two years and it was also more relevant to modern Android handsets. The data we used of these two years 
covers 14,432 out of 20,169 total bug records and 46,806 out of 67,730
total bug comments from the whole dataset. Related to these bug and comment records, there
are 30,969 people who have either reported a bug or made comments on a
bug.

The bug and comment records are grouped into 30-day windows sliding by
1 day. 
%The strategy of taking 30 days as a window and sliding by 1 day
%would be discussed in detail in Section \ref{methodology}. 
We extracted 673 windows in total from the bug reports during year 2010 and 2011.

%  The 
% ``project'' \ and the ``target'' \ schema record all the submitted files from
% the developers, from which the actual behaviours could be identified.

\subsection{Windowing Bug Reports and Extracting Social Networks \\Methodology}
We windowed the data and constructed networks
that indicated the relations among the participants within each specific
time window. For each window, we calculated the betweenness centrality of each
participant and we plotted the centrality values per participant in a visualization. The steps of our methodology are
explained as following:

\textbf{Step 1: Pruning the data.} We pruned the records of the
reporters and commenters into a pure name format, which are originally
recorded in semi-anonymous email formats in the XML repository dump. 
For example, given the original email address which is represented by
\textquotedblleft mathias....@gmail.com\textquotedblright, we truncate
the string starting from \textquotedblleft ....\textquotedblright \
and keep the front part \textquotedblleft mathias\textquotedblright \
at the beginning as the name of the reporter or commenter. 
This strategy could lead to name aliasing problem, especially for
common names or email addresses starting at just a simple letter like
\textquotedblleft e....@gmail.com\textquotedblright.
 Although algorithms have been provided to reduce the extent of the problem, \cite{ACM:robles}, \cite{MSR:christ}, \cite{Elsevier:goeminne}, it is difficult or even impossible to eliminate the influence from
 this data quality issue. When applied to other repositories that do not anonymize this would be less of a problem. Hence, we focus  on participants whose names are less common and less ambiguous in our study.

\textbf{Step 2: Windowing the records.} We windowed the data into
periods of 30 days with a 29-day overlap. 
30 days was chosen as a window size because it is smaller than the
periods between a major and minor release, it is similar to a month of
work, but long enough to contain the resolution of multiple bugs.
We have compared sliding by 1 day with our previous result of sliding by 7
days, 1 day sliding produces gradual and smoother transitions of
centrality.

\textbf{Step 3: Establishing the network.} We made a tool to perform the SNA with sliding
windows. The tool is implemented in Java and built on top of the JUNG Graph Framework, that converted bug
reports and bug comment records within a window to a social network graph.

The nodes of these networks represent participants who have either
reported some bugs or made comments on bugs. The edges represent
connections between two nodes. All the edges are weighted.  
For a bug within a selected time window, whenever a person
makes a comment on this bug, the edge between the bug commenter and
the bug reporter would get weight plus one, as well as the edges
linking to the participants who previously made comments on this
bug. Bug reports or comments in different windows would have separate
network graphs depending on the activity of their reporters or
commenters. An example in Figure 1 indicates how the weighted network
graph is built.

\textbf{Step 4: Calculating the centrality.} We calculated the
betweenness centrality using JUNG, and normalized the centrality with the number of node pairs, as in Equation (\ref{normalizedb}). We then get a list of all the bug participants and
their betweenness centrality values for the total 673 overlapping
windows.

\textbf{Step 5: Removing irrelevant participants.} We removed the
participants with betweenness centrality value 0, who might have
either reported a bug/bugs with no comments, or made the only comment
on a bug so that no other participants are related. Afterwards, we get
1654 participants with betweenness centrality value larger than 0, out of the 30969 in total.

\textbf{Step 6: Generating the analysis graph.} The activity
of each bug participant is represented by a 673 dimensional vector representing their betweenness per window. Each element of the vector indicates the betweenness centrality value extracted
from the graph, which is generated from the window for that
specific time period (in our case, the specific time period is 30 days
starting from the date of the window start point). Then we clustered
all the vectors by using K-means ($k=100$) with cosine distance to 100
clusters. Finally, we plotted the results, as shown in Figure 2 to
visualize the clustered data so that we could easily analyze
our results.

\subsection{Validation using the Android Release History and the Git} 
In addition to the methodology of mining the Android
bug repository, we made use of the git version control repository and
inspected the release history highlights to validate the purpose behind the
clusters and patterns we observed.
We looked into the participants who contributed to the git repository in order to find their areas of expertise and validate our analysis conclusion about how the community participants
act in accordance with the project development.

The types of files modified and the corresponding projects are highly
correlated with the specialization of those who commit changes. For
instance, if a developer always submits kernel related code
files, he is more likely to be specialized in kernel techniques. Types of files
include document files, test files, source files, etc; dictionary paths of files usually indicate what projects the files belong to. 
We manually identified the participants' areas of expertise by
observing the \emph{project} and the
\emph{target} for all of their commits (such as source code or
documentation).
To give a specific example, if there were commits from a developer,
\texttt{Mr.Guilfoyle}, on the target file
\texttt{media/java/android/media/Ringtone.java} under the project
\texttt{platform\_frameworks\_base}; then, we would suggest that \texttt{Mr.Guilfoyle} 
likely has some specialized knowledge about the platform's ringtone. 
Thus this is how we derive participant expertise~\cite{ICSEsocio:meneely}.

Also, we could further relate their expertise to their centrality
patterns. The Android release history could, on the other hand, help
to relate the release highlights to participants central behaviour during
that release. Further validation is discussed in Section
\ref{validation}.

\begin{figure}[!t]
\centerline{\includegraphics[width=3.56in]{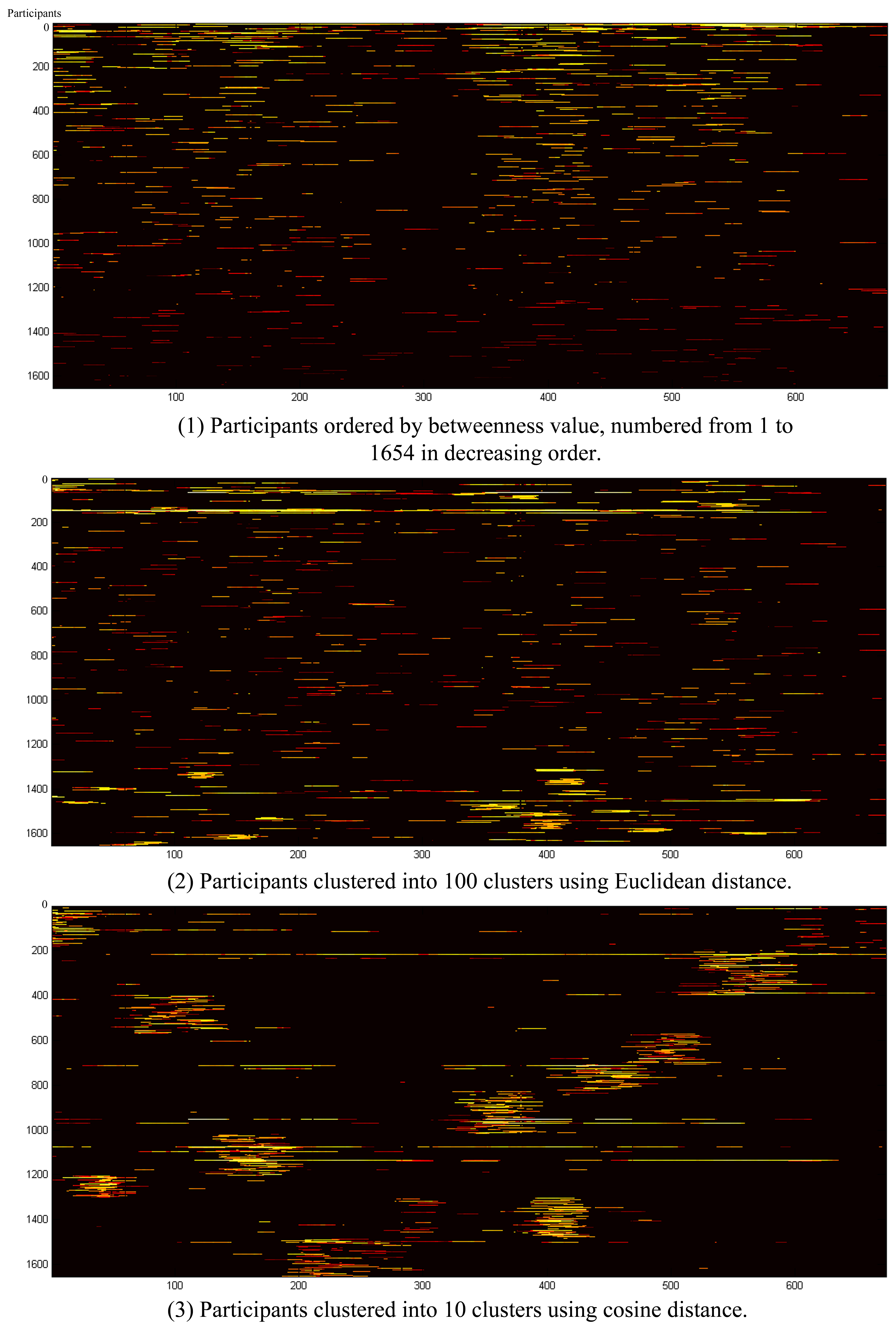}
\label{cluster_comp}}
\caption{Betweenness centrality along time line. Participants on the y-axes are ordered differently by betweenness values or various clusterings. }
\end{figure}

\section{Results and Analysis}
% describe the figure. 
% take up two columns
\label{results}

We study the results shown in Figure 2. Each horizontal line represents the 673 betweenness
centrality values for the selected bug participant during year 2010 and
2011. In total, we have 1654 bug participants. By studying these results, we answered the
following questions:

\begin{figure}[!t]
\centerline{\includegraphics[width=4.1in, height = 3.8cm]{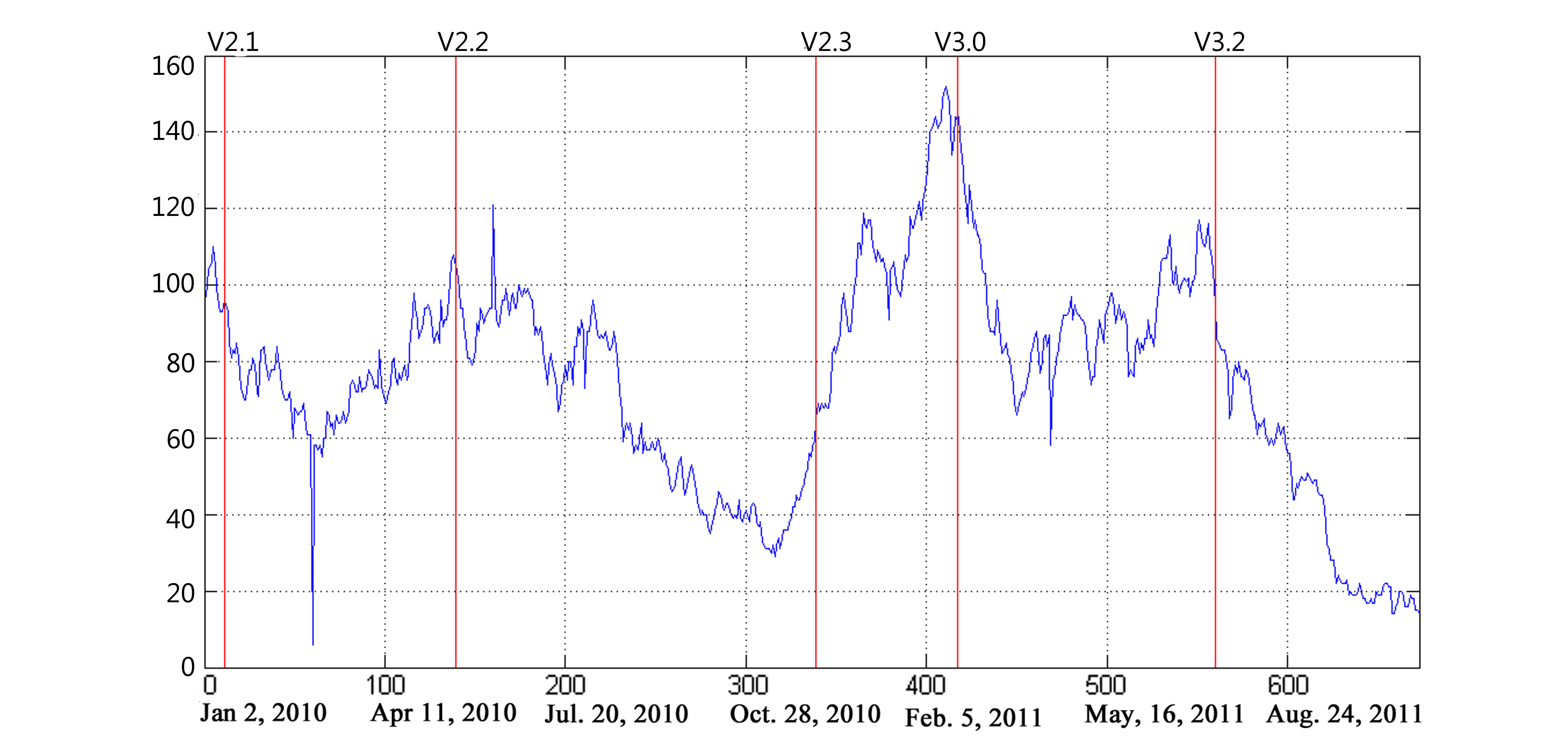}
\label{people}}
\caption{Number of active participants across time.}
\end{figure}

\begin{figure}[!t]
\centerline{\includegraphics[width=4.1in, height = 3.8cm]{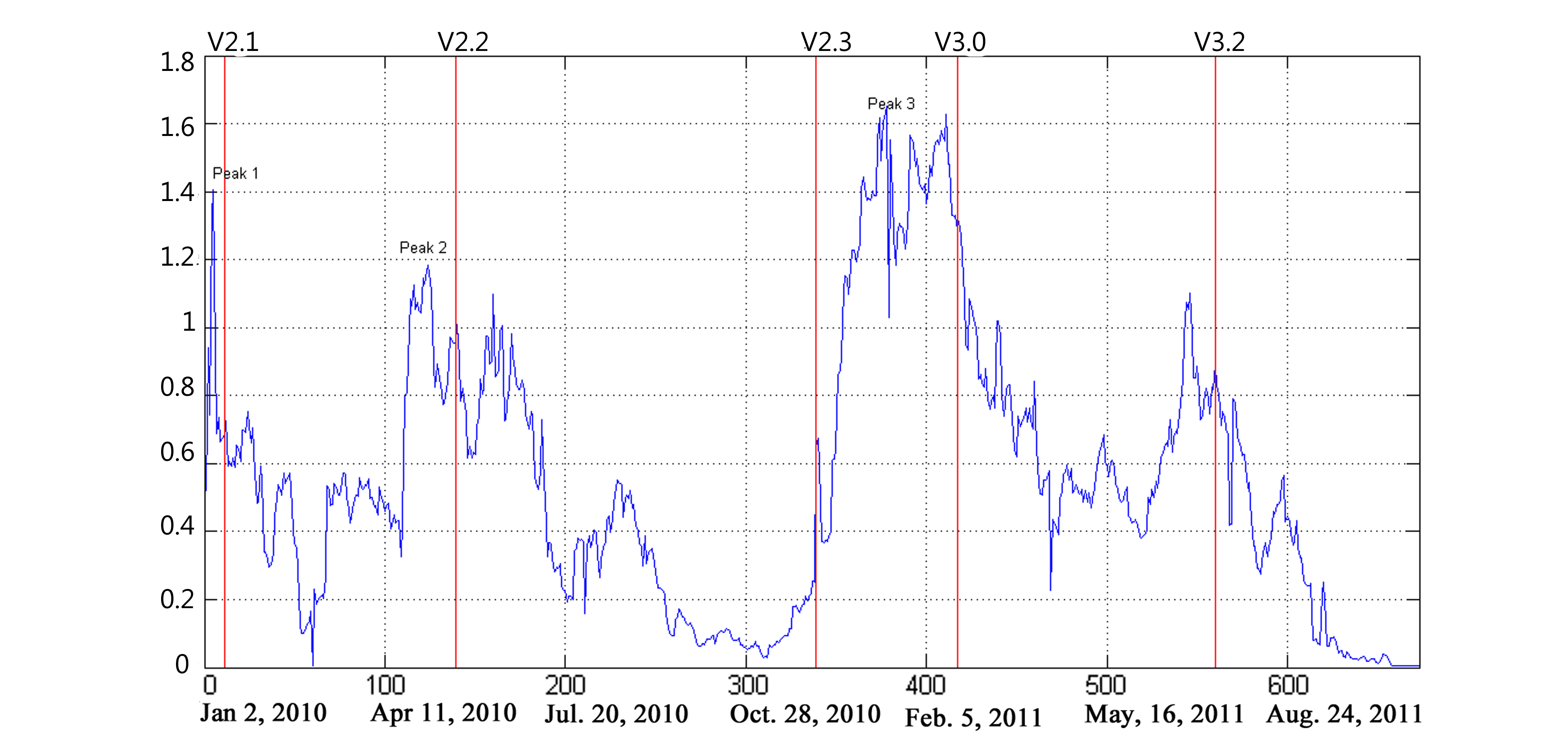}
\label{betweenness}}
\caption{Sum of betweenness centrality of participants across time.}
\end{figure}

\subsection{Global Analysis}

\noindent\textbf{RQ1. How does the number of active bug participants change over time? Why?}

To give an overview, we compared the interaction of bug participants
between January, 2010 and December, 2011, and found that the
interaction among participants in the Android bug community in 2011 was similar
to the interaction of participants in 2010 but more frequent, as we can see in Figure 2. One
``gap" occurs around window 300,
which we will explain in the next Local Analysis subsection. 

Correspondingly in Figure 4, that we counted the number of participants with betweenness centrality value larger than 0 within each window, the number of active participants
during 2011 is slightly larger than that of 2010. Figure 5 shows the sum of betweenness values along the two years' time line, we can see that the trend is very similar to that of the number of active participants in Figure 4. This also suggests that the betweenness centrality reflects the interaction among participants.

Moreover, a possible reason for the changes of the number of active bug participants
and the betweenness centrality values is that around major or minor
releases of SDKs, API fixes or improvements, participants seem to become more
active in bug reporting, discussing and fixing
activities. Also, during these time periods, bugs are more likely to
be discovered and reported. 
Perhaps the pressure of the release is 
causing developers to address outstanding bugs more than usual.
% AH: You can't just say this stuff, if you don't know you gotta admit it
% would probably work longer and harder than usual with more
%effort and passion, which promotes the discovery of more bugs and
%problems of the project. 
%And when the release date is approaching,
%they have to work hard to fix these bugs. 
After a
release, users also take part in the activity of
discovering the bugs and problems so that in this case both users and
developers would like to discuss the bugs. 
%The bug community becomes
%lively. 

\noindent\textbf{RQ2. How does the betweenness centrality of a participant change over time? What are the reasons when they have a certain activity pattern?}

Observing the continuity of betweenness centrality in Figure 2, some
participants have kept active during the entire two years, and
correspondingly they have a very continuous and bright line. For
participants of this type, there are a few possible explanations. 
% XXX IN THE PRESENTATION YOU NAMED NAMES. STOP GUESSING. TELL US IF
% THIS IS THE CASE
First, our conjecture is that these participants are professional developers who
belong to the core development team so that what they reported are
more important issues which attract more participants to discuss and fix
them. 
% XXX VERIFY THIS SENTENCE IS IT TRUE OR NOT?
Their identities of being professional developers will be discussed in Section \ref{activityvalid}.
% XXX YOU SAID THIS IN THE PRESENTATION AND HAD EXAMPLES DID YOU LOOK
% AT THIS OR NOT

Second, some of these participants are of high community status or
expertise, and they might
supervise and guide the development of the 
project. For example, when we validated, we did find one developer, romainguy, who has experiences on almost every component relevant to platforms so that he can be considered to be an expert.
%They would like to report big issues and will give expert
%comments on bugs. 
% XXX PLEASE GIVE AN EXAMPLE WHERE DID THE VALIDATION GO
Developers related to these continuous lines are listed in Table \ref{continuous_project}, and we will further discuss and validate on them in Section \ref{activityvalid}.

However, in most cases, participants' betweenness values are
highly variant, as observed in Figure 2. 
To investigate the variation in betweenness values over time, we
decided to count the number of times that a user experienced a range of consecutive windows in
which the user had non-zero betweenness.
% We counted the
% number of occurrences of the continuous betweenness values which are
% greater than 0. 
% It means that we would count 1 if the betweenness
% value is greater than 0 for the first time and take the next values
% together as the first continuous occurrence until we meet 0. 
% Then the
% second time we meet value greater than 0, we would count 2 and take
% the next values together as the second continuous occurrence until
% another 0 occurs. 
 
Participants with a count of distinct ranges
 greater than 1 would be \emph{phasers} who periodically participate within
 the Android bug
community. Here phasers are those who
phase into centrality and later out of it.
These randomly phasing participants (phasers) are very likely to
acquire less expertise or have lower community status in their
community, than those with continuous high centrality. 
Phasers
might be interested in limited topics and only central and active
during the appearance of bugs relevant to those topics.
Participants who only had 1
distinct range of betweenness are considered to be participants who
only appeared once, and are probably users. We validate the roles these participants play in Section \ref{activityvalid}.

To summarize, among the 1654 participants with betweenness values larger than 0, we
analyzed their centrality patterns and divide them into three
categories: 
1) participants appeared only once with a betweeness greater than 0 (71 out of 1654
participants),
2) participants recurred periodically (1575 participants)
and 3) participants who are central along the entire project
history (8 participants).

\subsection{Local Analysis}
\label{local}

\noindent\textbf{RQ3. Are there special time ranges during which participants are more/less active or central than normal? Why?}

By inspecting the Android release history highlights, we found that the v2.1
SDK was released on 12 January 2010, which corresponds to the first peak
value in Figure 5. Android v2.2 SDK was released on 20 May 2010 and this
corresponds to peak 2. From Dec. 2010 to the beginning of Mar. 2011,
several minor updates were released and on 22 Feb. 2011, one major
update v3.0 SDK was released. These releases explain the summit, i.e.,
peak 3, in  Figure 5. This is correlated with more participation at
the same time.

In addition, during the first obvious ``gap'', which covers the time from
October 2010 to the end of 2010 (around window 300), the social network during this time period is almost inactive
and even ``quiet''. There were fewer
releases during the ``gap". 

The other low value showing up in the end of Figure 5 results from
the fact that there are no bug reports recorded (right tail censoring) in the given dataset.
This piece of data is still meaningful because it contains comments belonging to bug reports several weeks or months before. The betweenness value is thus simply calculated by the comments here. 

\noindent\textbf{RQ4. What are the possible scenarios for a very sharp change of the participants' centrality? Why?}

Considering individual participants, almost all of them has
experienced centrality oscillations. In addition, some participants
tend to become active and core members during the same time
period and then they fade away together.

We suspect that the phasers tend to be interested in one or several
categories of problems so that they appear only along with the
occurrence of these issues. They take part in activities related to
the bugs or technical issues and become inactive after the problems are
solved. Or in the case when they are working on a project, they would become inactive when the projects are finished. 
As showed in Figure 2, the participants' tend to get clustered together around important releases, which supports that the phasers are working along with projects or related issues.
Meanwhile, by observing the clustered participants of their
activity patterns in Figure 2, we suspect that the phasers that show
up densely together could be interested in similar categories of
topics. 
This assumption is validated in Section
\ref{clustervalid}.

%XXX again you don't answer a research question by making stuff up
% relate this to your validation or remove it.
% If it can't be validated don't bother saying it as an answer to a question.
%Meanwhile, other possible explanations for the very sharp change in
%the participants' centrality could be: 1) the participants are pure
%users; they showed up merely to report bugs they found. These bugs
%attract other users or developers to take part in the discussion,
%which would increase the betweenness centrality of the bug reporters,
%i.e., the users, for a very short time period. 2) The participants are
%developers and are on a temporary leave, a vacation for
%example. 
%3) The project is just
%finished or discarded so that both users and developers would not
%discuss anything about it any more. This group of participants could be
%verified by looking for those who do not own any submission and then
%regarded as pure users. 
%The third explanation would be easily
%substantiated by inspecting Android version release history. However,
%the second explanation is hard or even impossible to validate with the
%very limited information we could get from the repositories and the
%release history. Participants should be contacted in person for such
%validation, which has not been done in our study.
%XXX GRRR Answer the question! Stop making stuff up. Link it to evidence.

%XXX This section needs to be integrated with the research question
%answers. I don't have the time to do this. 

\section{Validation}
\label{validation}
% This section needs to be modified!!!
% two parts: 1. assumption, take up 2 columns 2. validation, take up 3 columns
% validation: 1. clusters 2. phasers
We made use of the git repository and inspected the
release history to validate our answers to the research questions in the previous section.
For RQ1, it could only get answered based on assumption and the number of active participants across time as we counted in Figure 4, but not thoroughly validated. RQ3 is intuitively answered when we match the betweenness distribution with the release history by time, and no further validation is needed. For RQ 2 and RQ4, we have made a detailed validation in this section.

\subsection{Activity pattern validation}
\label{activityvalid}
From the mining results, among the 1654 participants with betweenness values larger than 0, we notice that there is a small group of
participants who have been central for most of our analysis period (8 participants out of the 1654); another
relatively larger group appear without any recurrence (71 participants out of the 1654); the majority
would periodically become central in their community (1575 participants out of the 1654). Based on the
three activity patterns proposed in RQ2, we confirmed many of our previous suspicions:

%XXX MENTION WHICH RQ IS RELEVANT
\subsubsection{Participants that appeared only once tend to be pure users}

We look into the git repository to find the files
submitted by the 71 participants who have appeared only once in the
bug community. We found that only 7 of them have ever committed a
change, which means that these 7 are developers rather
than pure Android users. The rest do not have commits in the version
control system. This verifies our assumption that participants
appeared only once in the bug community would more likely to be pure
users, as introduced in RQ2.

%XXX MENTION WHICH RQ IS RELEVANT
\subsubsection{Participants showed up periodically should be a combination of users and developers}
% y=742-704 + 1
% x=474 - 430 + 1
% z = 1447 - 1418 + 1
% y = 1584 - 1543 + 1
% u = 1584 - 1543 + 1
% y=742-704 + 1
% y
% 39
% u
% 42
% u+y+x+z
% 156
% (5+13+9+7)/(u+y+x+z)
% .21794871794871794871

Periodically appearing participants are the majority and we call them 
phasers. 
% We sampled 4 continuous portions of them (continuous
% means they are continuously numbered in the clustered results):
% participants 430 to 474, 704 to 742, 1418 to 1447, and 1543 to
% 1584. For each of the four groups, the number of developers are 5, 13,
% 9, 7 respectively. They take up 17\%, 32\%, 20\%, and 18\% of the
% total number of participants in each group. 
Based on the methodology in Section \ref{methodology}, we looked into the commit history in the git repository in order to verify the expertise of phasers. With as many as 1575 participants, we sampled 156 participants. $21.8\%$ of the sampled participants were developer phasers, who have submitted changes.
We studied the expertise of the developer phasers from this
sample. All except two of them have worked on specialized tasks that
implied some specific kind of expertise or specialization.
The rest $78.2\%$ have never
submitted files to the development community. They are probably users
of Android. Thus, phasers consist of both users and
developers. This answers to our assumption of the phasers' role in RQ2.

\begin{table}[!t]
%% increase table row spacing, adjust to taste
% if using array.sty, it might be a good idea to tweak the value of
% \extrarowheight as needed to properly center the text within the cells
\caption{5 continuously central participants who have submitted changes to the git.}
\label{continuous_project}
\centering
%% Some packages, such as MDW tools, offer better commands for making tables
%% than the plain LaTeX2e tabular which is used here.
\begin{tabular}{|p{0.13\linewidth}|p{0.13\linewidth}|p{0.62\linewidth}|}
\hline
Participant & \#Submitted-\_changes & Related Project \\
\hline
fadden & 1259 & device\_samsung\_crespo, platform(bionic, build, dalvik, etc.) \\
\hline
xav & 3501 & platform(frameworks\_base, build, external\_bouncycastle, etc.), device\_sample, \\
\hline
mbligh & 80 & kernel(common, experimental, linux-2.6, msm, omap, qemu, samsung, tegra) \\
\hline
ralf (Ralf.-Hildebrandt) & 665 & kernel(common, experimental, linux-2.6, dalvik, external\_libpng, sdk,system\_core, etc.) \\
\hline
romainguy & 1455 & device\_htc\_passion, device\_samsung\_crespo, platform(build, cts, dalvik, development, external\_bouncycastle, libcore, ndk, apps(AccountsAndSyncSettings, AlarmClock,
Bluetooth, Browser, Calculator, etc.),
inputmethods(LatinIME, iOpenWnn, PinyinIME, CalendarProvider),
providers(DownloadProvider, GoogleSubscribedFeedsProvider), wallpapers(Basic, LivePicker,
MagicSmoke, MusicVisualization), prebuilt, sdk, system\_core) \\
\hline
\end{tabular}
\end{table}

%XXX MENTION WHICH RQ IS RELEVANT

\subsubsection{Participants who were continuously central for a long time period could have multiple areas of expertise}

5 out of the 8 participants in this group have submissions in the
git.
% XXX What did you mean here VVV this makes no sense whatosever
% and they take up to 62.5\%, which is significantly
% larger than the previous two groups.
We extracted the projects these 5 participants have submitted changes
to, as listed in Table \ref{continuous_project}. (On the forth row, \texttt{ralf} and \texttt{Ralf.Hildebrandt} are email alias of the same person, as we observed that the author\_name attributes are the same for the two email alias.)

Firstly, considering the number of changes they made, all of them
except \texttt{mbligh} have more than 500 commits within the
git,
which means that they are quite active in Android
development community. This supports that they are
experts or advanced developers since more submissions indicates a
 broader range knowledge about the related techniques.
% If you don't know, don't say it
%, especially
%in such volunteered open source development community.

Moreover, \texttt{fadden}, \texttt{xav}, and \texttt{romainguy} are
all working on the platform layer,
which includes build, dalvik, development, framework base, libcore,
sdk, etc. All of their areas of expertise are related to the platform layer or
system core layer.
% and should be important along the entire
%development. 
%In this case, continuously central participants should
%have experiences on the system core related topics. 
%This is an
%interesting point beyond our hypothesis.

The participant \texttt{romainguy} has experiences modifying almost every
component relevant to platforms, including both the apps and the core, and
hence should be considered as Android platform development leader.

Furthermore, when investigating these continuous lines we found some
participants were Google employees, for example, two developers with alias \texttt{mbligh} and \texttt{romainguy}.
Their email account recorded in the git repository is from the \textquotedblleft google.com\textquotedblright \ domain, and moreover, when we googled them, they are indeed introduced as software engineers at Google. 

To summarize, this subsection demonstrates
that three different centrality patterns correspond to
participants of three categories, which supports our analysis hypothesis about activity patterns
in Section \ref{results}.

\begin{table}[!t]
\centering
\caption{5 clusters we have chosen, out of a total number of 21.}
\begin{tabular}{|c|c|}
\hline
Cluster & Time  \\
\hline
1 & May 16, 2010 - Jun. 24, 2010 \\
\hline
2 & Jun. 2, 2010 - Jul. 24, 2010  \\
\hline
3 & Jan. 13, 2011 - Mar. 3, 2011 \\
\hline
4 & Dec. 3, 2010 - Jan. 31, 2010 \\
\hline
5 & Feb. 4, 2011 - May 1, 2011 \\
\hline
\end{tabular}
\label{cluster_list}
\end{table}

\begin{table}[!t]
%% increase table row spacing, adjust to taste
\renewcommand{\arraystretch}{1.3}
% if using array.sty, it might be a good idea to tweak the value of
% \extrarowheight as needed to properly center the text within the cells
\caption{Participants and their areas of expertise in cluster No. 4}
\label{cluster_no4}
\centering
%% Some packages, such as MDW tools, offer better commands for making tables
%% than the plain LaTeX2e tabular which is used here.
\begin{tabular}{|c|c|c|}
\hline
ID & Name & Areas Of Expertise \\
\hline
1 & charles & kernel - sound, kernel\_linux-2.6 \\
\hline
2 & jasta00 & ringtone, media \\
\hline
3 & kristoff & driver(net, video, serial, input) \\
\hline
4 & rik(rik.bobbaers) & kernel\_linux-2.6(mlock) \\
\hline
5 & rik(rikard.p.olsson) & kernel\_linux-2.6(arm) \\
\hline
6 & rik(riku.voipio) & kernel\_linux-2.6(arm), driver \\
\hline
7 & snp & platform sdk(eclipse plugin) \\
\hline
\end{tabular}
\end{table}

\begin{table*}[!t]
%% increase table row spacing, adjust to taste
% if using array.sty, it might be a good idea to tweak the value of
% \extrarowheight as needed to properly center the text within the cells
\caption{Participants' common areas of expertise of each cluster. Participants number is counted as the number of participants within each cluster who has ever submitted a change and appeared in the git, ie., developers.}
\label{cluster_topic}
\centering
%% Some packages, such as MDW tools, offer better commands for making tables
%% than the plain LaTeX2e tabular which is used here.
\begin{tabular}{|p{0.1\linewidth}|p{0.13\linewidth}|p{0.64\linewidth}|}
\hline
Cluster & Participants Number & Areas Of Expertise \\
\hline
1 & 5 & netfilter, driver(video), tests, MIPS \\
\hline
2 & 13 & driver(usb, wireless, mouse), sound, net, i386, performance(tools), input methods \\
\hline
3 & 9 & sound, driver, frameworks\_base, tests, platform, kernel \\
\hline
4 & 7 & sound, media, kernel\_linux-2.6, driver, platform sdk, kernel video/serial \\
\hline
5 & 63 & net(bluethooth, net driver, ipv$x$, kernel\_linux-2.6), driver(dvd, media, usb, gpu, net), ia64, sound, tests \\
\hline
\end{tabular}
\end{table*}

\begin{table*}[!t]
%% increase table row spacing, adjust to taste
% if using array.sty, it might be a good idea to tweak the value of
% \extrarowheight as needed to properly center the text within the cells
\caption{Highlights of identified clusters from Figure 2}
\label{release}
\centering
%% Some packages, such as MDW tools, offer better commands for making tables
%% than the plain LaTeX2e tabular which is used here.
%lp{3in}
\begin{tabular}{|p{0.08\linewidth}|p{0.1\linewidth}|p{0.58\linewidth}|p{0.08\linewidth}|}
\hline
Release & Time & Highlights & Related cluster\\
\hline
v2.2 & May 20, 2010 & camera and gallery, portable wifi, multiple keyword language, performance(general, browser), media framework, Bluetooth, kernel upgrade, APIs(media, camera, graphis, data backup, device administrator, UI framework) & 1 \\
\hline
v2.2.1 & Jan. 18, 2011 & bug fixes(one is about root and unroot), security updates, performance improvements & 3\\
\hline
v2.2.2 & Jan. 22, 2011 & fixed minor bugs, including SMS routing issues & 3\\
\hline
v2.3 & Dec. 6, 2010 & UI refinements, faster text input, power management, NFC, multiple cameras, download management, new multimedia, new developer features(gaming, communication, multimedia, garbage collector, event distribution, video driver, input, native access-audio, graphics, storage, development), linux kernel upgrade to 2.6.36, Dalvik runtime, mixable audio effects & 4\\
\hline
v2.3.3 & Feb. 9, 2011 & NFC, Bluetooth, Graphics, media, framework, speech recognition, voice search, API(identifier, build-in app, locales), emulator skins & 5 \\
\hline
v3.0 & Feb. 22, 2011 & UI design for tables, redesigned keyboard, improved text selection, copy and pase, connectivity options(USB, WIFI, media, keyboard, bluetooth), apps update, browser, camera and gallery, contacts, email, development support & 5 \\
\hline
\end{tabular}
\end{table*}

%XXX MENTION WHICH RQ IS RELEVANT
\subsection{Cluster validation}
\label{clustervalid}
As we have discussed above, participants are more active around
important releases. Moreover, we can observe from Figure 2 that
participants' centrality distributions tend to form into groups or
clusters, that often
are found around the releases.
Participants belonging to the same group become central during
the same time periods and then fade away together.

We labeled 21 visible clusters from Figure 2 and looked into five of them which are located more around releases. 
The five clusters we chose are listed in Table \ref{cluster_list}.

We extract changes submitted by the members of each cluster from the
Android git. (For those who do not have records in
the git, we regard them as pure users and do not
consider them in this case). After inspecting their submissions, we
would get an idea about what kind of tasks they have been mostly
working on.  
Based on release history and the commit logs we found 
that these clusters tend to be coherent efforts undertaken by multiple kinds of participants.
%XXX I have no idea what RQ you are talking about. PLEASE SPECIFY
% WHICH HYPOTHESIS

%We have found the following facts that prove our hypothesis:

%XXX MENTION WHICH RQ IS RELEVANT

\subsubsection{Participants clustered together share similar areas of expertise and tasks}

Our analysis in Section \ref{results} shows that the phasers that show
up densely together could be interested in similar categories of
topics or working on tasks related to the same area.

As described in Section \ref{methodology}, we extract the targets and
project names from the git for each member appeared
within the cluster. The areas of expertise could be inferred by the contents
of the targets and the topics of the projects. We summarized the
areas of expertise of participant clusters (from Figure 2)
 in Table \ref{cluster_topic}.

% An example of a floating table. Note that, for IEEE style tables, the 
% \caption command should come BEFORE the table. Table text will default to
% \footnotesize as IEEE normally uses this smaller font for tables.
% The \label must come after \caption as always.

Inspecting the areas of expertise, we find that each cluster has their own
topics, which are relatively different from each other. Also, the topics of
each cluster are concentrated to specific layers of Android's architecture.
For example, cluster No.1 covers techniques about net filters,
drivers, tests, and MIPS, while cluster No.2 is about drivers for
connecting devices (usb, wireless, and mouse), net, processor, and
input. 
It is easy to tell that participants of these two clusters are
working on different tasks. 
The other clusters could lead to the same
conclusion. 
Thus we conclude that clusters often exist around a topic.

% AH: not a clear or useful point
% if we look into one of the clusters, we can see that
% sub-groups exist in each cluster, and each sub-group works on one
% common topic. 

Take cluster No.4 as an example. There are 7 developers
contained in this cluster, as listed in Table \ref{cluster_no4}.
It can be observed that work of participants in this cluster
could be generally divided into two groups: one is about the Linux 2.6
based kernel, another is related to multimedia. \texttt{Charles}, \texttt{rik.bobbaers}, \texttt{rikard.p.olsson}, and \texttt{riku.voipio} (the pruned bug reporter alias \texttt{rik} is related to three developers in the git and we look them all; this issue would be discussed in Section \ref{limitation}) are all modifying the Linux 2.6
kernel. \texttt{Charles}, \texttt{jasta00}, and \texttt{kristoff} are working on
multimedia topic, which includes sound, video drivers, and
ringtone. 
% XXX DONT MAKE STUFF UP
%Besides, we even guess that these two areas of expertise may have
%strong connection so that those who work on the kernel are likely to
%work on the media related part. 
% For example, the projects Charles Chin
% is working on are all about the kernel, with sound as the target,
% which means he is working on the sound related kernel, and it is the
% same with Kristoffer.
% l

% XXX did you look at other clusters?
When we look into other clusters, we get similar conclusions.
Thus, from the observation and analysis above, we can conclude that
participants with similar centrality patterns often share similar
areas of expertise and tasks. This validates our assumption about the phasers being clustered on specific techniques in RQ4.

%XXX MENTION WHICH RQ IS RELEVANT
\subsubsection{Clusters' working areas of expertise are in accordance with the release contents along the time line}

When observing the Android release history, we concluded
that the overall betweenness centrality becomes higher around
releases, and more active participants appear around important
releases, at least according to Figure 4 and Figure 5.

In addition, when taking participants' areas of expertise into consideration,
we find that the release highlights are in accordance with the
areas of expertise for members of each cluster. Table \ref{release} lists
releases and their corresponding clusters together with the
highlighted release contents.

Comparing the release contents and the cluster areas of expertise, these two
subjects are mostly matched on release topics and cluster's working
contents. For example, cluster No.4 covers from December 3, 2010 to
January 31, 2011, which occurs before release v2.3. Participants in
cluster No.4 have areas of expertise relevant to sound, media, and kernel-video, which
match the release contents of new multimedia, APIs for native audio,
and mixable audio effects in v2.3; 
%areas of expertise about video drivers,
%input and keyboards match the contents featured on video driver and the
%faster text input accordingly. 
We can also find that 4 out of 7
developers in cluster No.4 have worked on the kernel when 
the linux kernel was upgraded to 2.6.36 in Android v2.3.

Cluster No.3 was centered around the releases of v2.2.1 and v2.2.2
(January 18, 2011 and January 22, 2011 respectively). Release 2.2.1
contained security updates and performance improvement; participants
in cluster No.3 are specialized mostly on kernels or platforms.  
This occurs in cluster
No.1 and its corresponding release v2.2 as well.

Our conclusion is that participants' work is relevant to areas of expertise
associated with clusters, and at the same time, the clusters and participation tends
to be correlated with releases.
This further validates our answer to RQ4 that developers tend to work as groups on specific projects or issues they are specialized, and their centrality patterns are related to the occurrences of projects or issues.

\section{Limitations}
% take up 1 column
\label{limitation}

In this study we explicitly trust that the same account of email
addresses, i.e., the part before \textquotedblleft
@\textquotedblright, belongs to the same bug participant. With the
given semi-anonymous email addresses in Android bug repository, we
pruned the part starting from \textquotedblleft ....\textquotedblright
~and kept the front part as the names of bug participants. However, it
is possible that some common names share the same start string. For
example, \textquotedblleft Benjamin Franzke\textquotedblright,
\textquotedblleft Benjamin Tissoires\textquotedblright  ~and
\textquotedblleft Benjamin Romer\textquotedblright  ~have the same
first name. We cannot distinguish these names with the email address
\textquotedblleft Benjamin@XXX". Besides, some of the email addresses
start with a simple letter which is ambiguous identifying a person,
while we analyze the results without excluding such data.

% In addition, among the 1654 bug participants whose betweenness
% centrality has ever been larger than 0, more than 1500 of them are
% periodically appearing participants. We sampled 4 groups of participants
% since the results from these sampled data are representative to
% substantiate the assumption we have made for periodically appearing
% participants.

We validate our analysis based on the assumption that the types and projects of
submitted files reflect the areas of expertise that the developers are specialized
in. Hence, we tagged the participants with the techniques according to
their submitted files in the Android git. However,
there could be inconsistency between the techniques
and the submitted files.

Our manual inspection increased the validity of the results, but it
still relied on the authors judgment, interpretation and potential bias.

\section{Conclusion and Future Work}
% take up 1 column
\label{conclusion}

In this paper, we mined the Android bug repository and studied the
data of 2010 and 2011. We combined overlapping time windows with
social network analysis in order to analyze the participants
interactions within the Android bug repository, as part of the Android open source community.

We conducted
a thorough case study of the bug reporter activity within the
Android bug repository with our method. 
We analyzed the temporal evolution of the Android
bug reporting community both globally and locally.  
We found that most
minor or major releases lead to high betweenness centrality in
general. We found and explained sharp changes of participants' betweenness values and we inspected three activity patterns for the participants. Also, we found out that participants tend to get clustered into groups. Then, we
validated these results by manually inspecting the Android version control
system (git) and the Android release history highlights. We validated the three activity patterns of bug participants as well as their corresponding reasons. 
For participants who
were clustered in same groups in our plots, they showed interest
in a set of similar topics as we inspected in our validation.

Thus we conclude that by combining the SNA with sliding windows, we were able to find many local interactions that would be lost in a global analysis. The sliding windows make these local collaborations more visible, instead of drowning them out in a global analysis. In this case, we can get a more accurate knowledge about participants' working patterns as well as their group working. Furthermore, we validated our findings by inspecting other repositories to confirm that the local behaviour occurred and was of relevance. This work could be used by managers and researchers to produce project dashboards, and automated project status reports. 

Future work includes applying the approach in this paper to other open
source projects' repositories in order to improve its generality. We
want to further validate if our overlapping time windowing SNA plots
are trustworthy enough to depict the actual develop processes of
various projects.

\clearpage

%Include references using bst and bib files.
%\bibliographystyle{mybstfile} 
%{\small \bibliography{mybibfile}}

\bibliographystyle{ieeetr}
\bibliography{asg}

\end{document}